\documentclass[12pt]{iopart}
\usepackage{setstack}
\usepackage{verbatim}
\usepackage{bm}
\usepackage{graphicx}

\begin{document}

\newcommand{\ket}[1]{| #1 \rangle}
\newcommand{\bra}[1]{\langle #1 |}
\newcommand{\smallfrac}[2]{\mbox{$\frac{#1}{#2}$}}

\title{Radiative properties of a linear chain of coupled qubits}

\author{C J Mewton and Z Ficek}
\eads{\mailto{mewton@physics.uq.edu.au}, \mailto{ficek@physics.uq.edu.au}}
\address{Department of Physics, School of Physical Sciences,
The University of Queensland, Brisbane, Queensland, 4072 Australia}
\date{\today}

\begin{abstract}
We calculate the radiative properties for a linear dipole-coupled chain of qubits.  Using the explicit energy eigenstates of the system, we find the radiation patterns for spontaneous transitions from the one-photon eigenstates to the ground state of the system.  We show that depending on the excitation of a specific atom, the radiation tends to be focused either along or perpendicular to the chain.  We conclude with a derivation of the total decay rate of the one-photon eigenstates, and find the interesting result that for systems where the photon wavenumber is not much larger than the interatomic spacing, up to 94\% of the eigenstates are subradiant, that is, they decay significantly slower than a single atom in isolation.
\end{abstract}


\maketitle

\section{Introduction}

There is considerable interest in the study of the entanglement properties and radiative dynamics of systems that are comprised of qubits, where the number of qubits is known exactly, with the aim of using such systems for quantum information processing.  An understanding of the radiative properties of such a system is of great use since it provides information about decoherence and the radiation (transmission) channels in a multi-atom system, and how useful the spectrum of eigenstates are with respect to their persistence in time.  Such information could be useful for the study of controlled, decoherence-free information transfer through the chain.

The solution of a linear spin chain with a Heisenberg interaction was first given by Bethe \cite{bib:Bethe1931, bib:Yang1966}.  He analysed the properties of a linear spin chain in order to develop a theory of ferromagnetism.  The system was assumed to have periodic boundary conditions, and he introduced a diagonalisation ansatz which allowed the eigenstates of the system to be obtained.  The form of these eigenstates could be interpreted as ``spin waves'' travelling along the chain.  Using this solution, he was able to obtain the partition function for the system.  From the partition function, one can calculate many thermodynamic properties which describe the system macroscopically, such as the magnetisation \cite{bib:Katsura1962}.  Subsequently, a great deal of attention was paid by later authors to determining the energy eigenvalues, and in turn the partition function.  The ground state and one-photon excitations received considerable attention, e.g. in \cite{bib:Bethe1931, bib:Yang1966, bib:Katsura1962,  bib:Lieb1961, bib:Katsura1963}.

Research followed in Bethe's direction, in considering various types of linear spin chains and investigating their macroscopic properties.  The XY Heisenberg chain was solved exactly by Lieb \emph{et al.} \cite{bib:Lieb1961}, for example.  They also investigated correlations between spins, such as the long-range order.  The anisotropic XYZ model was treated by \cite{bib:Katsura1962, bib:Katsura1963}, who studied various macroscopic properties, including the free energy and susceptibility.  The partition function and the energy eigenvalues of the XY chain were also presented and were in agreement with \cite{bib:Lieb1961}.  The ground state energy for the XYZ model was later obtained in the general case in \cite{bib:Baxter1971}.

A number of analytical studies relevant to quantum information processing have been 
performed on systems composed of $N$ dipole-coupled atoms confined to fixed 
positions, with a particular interest in ring
arrangements, i.e. a small number of atoms arranged in a ring at equidistant positions \cite{bib:Freedhoff2004, bib:Rudolph2004, bib:Freedhoff1986b, bib:Hammer2004}.  Such forms require periodic boundary conditions and possess $N$-fold symmetry, thus making it easier to determine the energy eigenstates and energy levels.  There is an introduced complication, however: in a line, one can consider only nearest-neighbour couplings; in a ring, additional couplings beyond the nearest neighbour must be considered.  This clearly follows from the geometry of a ring as opposed to a line.  Refs \cite{bib:Freedhoff2004, bib:Rudolph2004, bib:Freedhoff1986b} analyse ring arrangements and develop a technique which can be applied to any $N$-atom ring system.  It is also demonstrated that it is important to consider all interactions, not just nearest neighbours.  As the solutions are not generalised for arbitrary $N$, a list of eigenstates are given for various values of $N$, but only up to the first three excitation levels.

The eigenstates corresponding to the elementary excitations of a ring and a ring with an atom at the centre are given in \cite{bib:Hammer2004}.  These states are used to study the associated decay rates.  Energy eigenstates and eigenvalues have also been obtained for a linear spin chain of alternating atoms in \cite{bib:Feldman2005}.  Many papers have appeared on various aspects of entanglement in linear chains of qubits \cite{bib:Arnesen2001, bib:Wang2001, bib:Hiesmayr2006, bib:Kofler2006, bib:Guehne2005}.

Recently, we have presented the analytic expressions for the energy eigenstates and energy eigenvalues for a linear chain of identical, two-level atoms (qubits) \cite{bib:Mewton2005}.  Such a system is mathematically equivalent to the Heisenberg XY model, which was analysed by Lieb \emph{et al.} \cite{bib:Lieb1961}.  Although they diagonalized a more general Hamiltonian, we will demonstrate that our restriction to a specific interaction model, and not presenting solutions in terms of the Jordan-Wigner transform makes for a more perspicuous presentation in that the eigenstate basis is obtained using a new method, and the basis used is directly related to what can be measured, i.e. the excitation states of the atoms.

A linear chain of atoms has been realized experimentally using ion trap techniques \cite{bib:Haffner2005}.   In these experiments the systems are controlled by individually interacting with the ions with focused laser light \cite{bib:Retzker2005, bib:Beige2000b}.  However, the experimental form does not make use of the dipole-dipole interaction.  Our approach, on the other hand, is to explore the dipole-dipole interaction in its potential to create multi-particle entanglement.  Moreover, it allows one to selectively control atoms through the collective excitations of the system rather than individually addressing atoms with focused laser beams.  To do this, however, we need to know how the system will interact with the electromagnetic field.

  In this paper, we use the state vectors given in our previous paper to explore the radiative properties of the system.  We are specifically interested in the angular distribution of the radiation emitted by states that lie within the first excitation manifold.  In addition, we will determine how many one-photon eigenstates are subradiant, i.e. exhibit reduced decay rates, and discuss its dependence on the number of atoms.  In quantum information processing, it is ideal that we have a \emph{decoherence-free subspace}.  Although in general most of the eigenstates in the system presented here decay after a finite time, most of the one-photon states are subradiant, which is also beneficial.

The paper is organized as follows.  We first look at state properties of the chain in Sec. \ref{sec:properties}.  In this section, we  summarize the main result of the previous paper in which we derived the energy eigenstates and eigenvalues of the chain.  However, we are now able to present the normalized states; the derivation of the normalisation constant is given in Sec. \ref{sec:ortho} on orthonormality relations.  Next, we derive a parameter which gives a measure of the symmetry of a particular state, of which there are two possibilities: symmetric or antisymmetric with respect to reflection through the middle of the chain.

In Sec. \ref{sec:eigendecay}, we derive the transition probability per unit solid angle $d \Gamma_g / d \Omega$ in a transition from a $M=1$ state to the ground state.  The analysis concludes with a demonstration that the results for a two-atom system which are known in the literature can be recovered from our present work.  We next explore in Sec. \ref{sec:singleAtomDecay} the decay of the system when it is known which atom is excited, and find that the transition probability can be expressed as a simple superposition of eigenstate transition probabilities.  In Sec. \ref{sec:totalDecayRate}, we look at the total decay rate of the one-photon eigenstates so that we can get an idea of the size of the subradiant subspace.

\section{General properties of the eigenstates}
\label{sec:properties}
Recently, we developed a new technique for finding the multi-quantum eigenstates 
in a linear chain of $N$ identical, equally spaced and confined to fixed
positions, two-level atoms each interacting only with its nearest neighbors 
through the dipole-dipole interaction \cite{bib:Mewton2005}.  The system we consider is governed by the Hamiltonian
\begin{eqnarray}
\hat{H} = \hat{H}_0 + \hat{V} = \hbar \omega_0 \sum_{i=1}^{N} \hat{S}_i^z
+\hbar \Omega \sum_{ \substack{ i,j=1 \cr |i-j|=1 }}^N
\hat{S}_i^+ \hat{S}_j^- ,
\label{eq:hamiltonian}
\end{eqnarray}
where $\hat{H}_0$ is the interaction-free Hamiltonian and 
$\hat{V}$ is the dipole-dipole interaction between the atoms.

In (\ref{eq:hamiltonian}), $\omega_0$ is the transition frequency 
of a two-level atom in isolation, $\hat{S}_i^z$ is the energy operator of 
the $i$-th atom, and $\hat{S}_i^+$, $\hat{S}_i^-$ are the raising and
lowering operators for the $i$-th atom, respectively.  
The dipole-dipole interaction parameter $\Omega \equiv \Omega(r_{ij})$
depends on the distance $\mathbf{a}$ between adjacent atoms in the linear 
chain and the polarization of the atomic dipole moments $\boldsymbol{\mu}$ relatively 
to the interatomic axis.

\begin{figure}
\begin{center}
\includegraphics{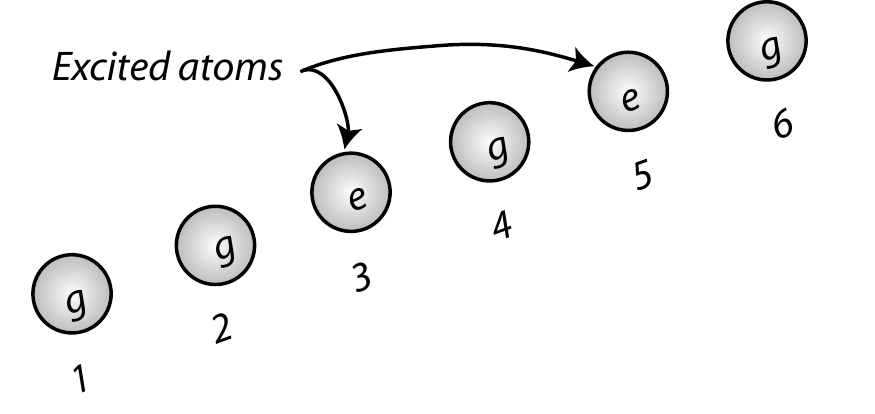}
\end{center}
\caption{As an example of our basis notation, the collective state shown in the figure for a 6 atom chain is represented by $\ket{3,5}$.}
\label{fig:notation}
\end{figure}

We can index the atoms in the
chain by the numbers $1$ to $N$; collective excitations the system can then
be represented using the ket $\ket{k_1, \ldots, k_M}$, where the non-zero
integers $k_1, \ldots, k_M$ denote the indices of the atoms which are in
their excited state:
\begin{equation}
\ket{k_1, \ldots, k_M} = \hat{S}_{k_1}^+ \cdots \hat{S}_{k_M}^+ \ket{0} ,
\end{equation}
where $\ket{0}$ denotes the ground state of the collective system, and is
nondegenerate.  A graphical example of this notation is given in figure \ref{fig:notation}.

As we have shown in \cite{bib:Mewton2005}, the unnormalized eigenstates for the system are given by
\begin{eqnarray}
&& \ket{\psi^{(M)}_{g_1 \cdots g_M}} = \sum_{k_1 < \cdots < k_M}
 C^{k_1 \cdots k_M}_{g_1 \cdots g_M} \ket{k_1, \ldots, k_M},
 \label{eq:energyVulgarEigenstates}
\end{eqnarray}
with corresponding eigenvalues
\begin{eqnarray}
E^{(M)}_{g_1 \cdots g_M} & = & M \hbar \omega_0
 + 2 \hbar \Omega \sum_{i=1}^M \cos (g_i \xi ),
\end{eqnarray}
where
\begin{equation}
C^{k_1 \cdots k_M}_{g_1 \cdots g_M} = \epsilon_{a \cdots d}
\sin ( g_a k_1 \xi ) \cdots \sin ( g_d k_M \xi ) ,
\end{equation}
and $\xi = \pi / (N+1)$.  The integer $1 \le M \le N$ denotes the number of excited atoms in the system.  The integers $g_1, \ldots, g_M$, where $1 \le g_1 < \cdots < g_M \le N$, label the eigenstates.  In the present paper, we will work with the normalized form of the eigenstates:
\begin{eqnarray}
&& \ket{\psi^{M}_{g_1 \cdots g_M}} = \sum_{k_1 < \cdots < k_M}
\! c^{k_1 \cdots k_M}_{g_1 \cdots g_M} \ket{k_1, \ldots, k_M},
 \label{eq:energyEigenstates}
\end{eqnarray}
where
\begin{equation}
c^{k_1 \cdots k_M}_{g_1 \cdots g_M} = \left( \frac{2 \xi}{\pi} \right)^{M/2} C^{k_1 \cdots k_M}_{g_1 \cdots g_M}.
\label{eq:norm_c}
\end{equation}
The derivation of the normalisation constant is given in Sec. \ref{sec:ortho} on orthonormality relations.  Note the slight change in notation -- unnormalized states, like those given in our previous paper, are specified with the $M$ parameter enclosed by parentheses, whereas normalized states do not have the parentheses.

\subsection{Orthonormality relations}
\label{sec:ortho}
It is clear that states with different excitation levels $M$ are orthogonal.  Here, we show that states with the same $M$ but different $g_i$ are also orthogonal.  Eigenvectors corresponding to \emph{different} eigenvalues of a Hermitian matrix are automatically orthogonal.  However, our some of our eigenstates are degenerate and hence it is useful to show that the eigenstates we use are orthogonal, no matter the multiplicity of their associated eigenvalues.  As a byproduct of this proof, we also obtain the normalization factor for an arbitrary eigenstate as shown in (\ref{eq:norm_c}).  Consider the expression
\begin{equation}
\langle \psi^{(M)}_{g'_1 \cdots g'_M} | \psi^{(M)}_{g_1 \cdots g_M} \rangle
= \sum_{k_1 < \cdots < k_M} 
C^{k_1 \cdots k_M}_{g'_1 \cdots g'_M}
C^{k_1 \cdots k_M}_{g_1 \cdots g_M}.
\end{equation}

Although the $C^{k_1 \cdots k_M}_{g_1 \cdots g_M}$ coefficients are antisymmetric with respect to the $k_i$, the product $C^{k_1 \cdots k_M}_{g'_1 \cdots g'_M} C^{k_1 \cdots k_M}_{g_1 \cdots g_M}$ is symmetric under such an interchange, for the two elements of the product reverse sign and hence the overall product does not change sign.  This means that such a permutation will not affect the inner product of the two states; we introduce on the right-hand side of the above equation a sum over all permutations of the $k_i$, which simply produces $M!$ copies of the inner product.  To maintain equality, we divide this new sum by the factor $M!$:
\begin{eqnarray}
&& \langle \psi^{(M)}_{g'_1 \cdots g'_M} | \psi^{(M)}_{g_1 \cdots g_M} \rangle
= 
\frac{1}{M!} \sum_{\sigma(k_1 \cdots k_M)} \sum_{k_1 < \cdots < k_M} 
C^{k_1 \cdots k_M}_{g'_1 \cdots g'_M}
C^{k_1 \cdots k_M}_{g_1 \cdots g_M}.
\end{eqnarray}

The double sum in the above expression can be simplified.  Permuting the $k_i$ merely implies that each $k_i$ can go beyond its range of $k_{i-1} < k_i < k_{i+1}$.  This means that we can relax the restriction in the second sum that $k_1 < \cdots < k_M$, and let each $k_i$ range from $1$ to $N$.  The only extra terms this new summation introduces are ones where two or more of the $k_i$ are equal.  However, the $C^{k_1 \cdots k_M}_{g_1 \cdots g_M}$ vanish in such cases, so equality is maintained in transforming the double summation into one where all the $k_i$ range from $1$ to $N$:
\begin{equation}
\langle \psi^{(M)}_{g'_1 \cdots g'_M} | \psi^{(M)}_{g_1 \cdots g_M} \rangle
= \frac{1}{M!} \sum_{k_1, \ldots, k_M = 1}^N 
C^{k_1 \cdots k_M}_{g'_1 \cdots g'_M}
C^{k_1 \cdots k_M}_{g_1 \cdots g_M}.
\label{eq:orthosum}
\end{equation}

Let us now consider the form of each term in this summation:
\begin{eqnarray}
\fl C^{k_1 \cdots k_M}_{g'_1 \cdots g'_M} C^{k_1 \cdots k_M}_{g_1 \cdots g_M}
& = &
\epsilon_{a' \cdots f'} \epsilon_{a \cdots f}
\sin g'_{a'} k_1 \xi \sin g_a k_1 \xi
\cdots \sin g'_{f'} k_M \xi \sin g_f k_M \xi .
\end{eqnarray}

In order the right-hand side of (\ref{eq:orthosum}) to not vanish when summing over the $k_i$, we must have $g'_{a'} = g_a, \ldots, g'_{f'} = g_f$.  This means that if the $g_i$ for one state are different when compared to another state, the inner product of those two states vanish.  Only one individual $g_i$ need be different in order to make the entire expression vanish.  We now consider the case where the $g_i$ for the two states are the same, so that we may extract a normalisation constant.  In the above expression, we see that each $k_i$ occurs twice in the form of two sine factors.  It is clear that in the case where the $g_i$ are identical for the two states considered, the only terms that contribute to a nonvanishing summation are those for which a particular $k_i$ is multiplied by two identical $g_i$ in the two respective sine factors.  A summation over each $k_i$ then produces a factor of $(N+1)/2$.  Since there are $M$ sine pairs, each nonvanishing summation evaluates to $(N+1)^M/2^M$.  However, given one such nonvanishing sum, there will exist other nonvanishing sums of the same form but with the $g_i$ in a different order.  This is due to the presence of the two Levi-Civita symbols.  There are $M!$ permutations of the $g_i$, so the summation will evaluate to $M! (N+1)^M/2^M$.  Therefore
\begin{equation}
\langle \psi^{(M)}_{g'_1 \cdots g'_M} | \psi^{(M)}_{g_1 \cdots g_M} \rangle
= \left( \frac{N+1}{2} \right)^M \delta_{g'_1 g_1} \cdots \delta_{g'_M g_M},
\end{equation}
where it is assumed, as always, that $g_1 < \cdots < g_M$.
We therefore deduce that the normalized eigenfunctions are
\begin{eqnarray}
&& \ket{\psi^{M}_{g_1 \cdots g_M}}
= \left( \frac{2}{N+1} \right)^{M/2}
\sum_{k_1 < \cdots < k_M}
C^{k_1 \cdots k_M}_{g_1 \cdots g_M} \ket{k_1, \ldots, k_M}. \quad
\end{eqnarray}

\subsection{Symmetry properties}

One would expect that the energy eigenstates of a linear chain would be symmetric or antisymmetric when the line is reflected through the centre of the line.  We now show that this is the case.  To reflect a state through the midpoint of the line, we replace all the $k_i$ in the coefficients $C^{k_1 \cdots k_M}_{g_1 \cdots g_M}$ by $k_i' = N+1 - k_i$.  We thus have
\begin{eqnarray}
C^{k'_1 \cdots k'_M}_{g_1 \cdots g_M} & = & \epsilon_{ab \cdots f} \sin g_1 k_a' \xi \cdots \sin g_M k_f' \xi \nonumber \\
& = & \epsilon_{ab \cdots f} \sin g_1 (N+1 - k_a) \xi 
\cdots \sin g_M (N+1 - k_f) \xi .
\end{eqnarray}
Next, we expand the sine terms as
\begin{eqnarray}
\sin g_i (N+1 - k_j) \xi & = & \sin g_i (N+1) \xi \cos g_i k_j \xi
 - \cos g_i (N+1) \xi \sin g_i k_j \xi \nonumber \\
& = & \sin g_i \pi \cos g_i k_j \xi - \cos g_i \pi \sin g_i k_j \xi \nonumber \\
& = & (-1)^{1+g_i} \sin g_i k_j \xi .
\end{eqnarray}
Hence we have
\begin{eqnarray}
C^{k'_1 \cdots k'_M}_{g_1 \cdots g_M} & = & \epsilon_{ab \cdots f} [(-1)^{1+g_1} \sin g_1 k_a \xi ] 
\cdots [(-1)^{1+g_M} \sin g_M k_f \xi ] \nonumber \\
& = & (-1)^{M + g_1 + \cdots + g_M} \epsilon_{ab \cdots f} \sin g_1 k_a \xi   \cdots \sin g_M k_f \xi \nonumber \\
& = & (-1)^{M + g_1 + \cdots + g_M} C^{k_1 \cdots k_M}_{g_1 \cdots g_M}
\end{eqnarray}
We must remember that $C^{k'_1 \cdots k'_M}_{g_1 \cdots g_M}$ is not in a canonical format, since the $k'_i$ are in the wrong order.  To bring them in the correct order (i.e. complete reversal) requires
\begin{equation}
\sum_{l=1}^{M-1} (M-l) = \frac{M(M-1)}{2}
\end{equation}
transpositions, hence there will be an extra factor of $(-1)^{M(M-1)/2}$.

If we think of an operator $R$ which reflects a given eigenstate through the midpoint of the atom chain, we can say that it operates in the following way:
\begin{equation}
R \ket{\psi^{(M)}_{g_1 \cdots g_M}} = (-1)^{M(M+1)/2 + g_1 + \cdots + g_M} \ket{\psi^{(M)}_{g_1 \cdots g_M}},
\end{equation}
where we have collected all the $M$ terms together.  
Thus if $(-1)^{M(M+1)/2 + g_1 + \cdots + g_M}$ equals $+1$, we say that the state is symmetric; if it equals $-1$, it is antisymmetric.  This agrees with the result in the first paper, which says that for $M=1$, $g_1$-even states are antisymmetric and hence do not decay in the small sample model, since $1+g_1$ is an odd number and hence $(-1)^{1+g_1} = -1$.

We would like to stress that the relationship between $g$ and symmetry can be misleading for the case $M=1$ due to an overloading of nomenclature.  The words ``symmetric'' and ``antisymmetric'' are often associated with the words ``even'' and ``odd,'' respectively, as is the case when discussing the parity of a function.  One may be mislead into thinking that the symmetry or antisymmetry of an eigenstate is related to whether its $g$-value is even or odd.  We must remember, however, that it is the even-$g$ states that are antisymmetric, i.e. ``odd states,'' and vice versa for odd-$g$ states.

\section{Spontaneous radiation patterns for $M=1$ to $M=0$ transitions}
\label{sec:eigendecay}

In this section, we study the directional properties of the radiation emitted by the chain in the far-field zone (in the far-field zone, photons emitted from either end of the chain will differ in phase, but not direction).  Directionality is an important factor as it determines the self-controlled directional properties of radiation that can be used to study the directional propagation of an excitation along the chain.  When an atom is excited by an external field, it may spontaneously decay to its ground state.  In the spontaneous decay of an isolated atom, the photon is usually lost and the atom is ultimately left in its ground state.  The radiation pattern produced by such an atom in free space is proportional to $|\boldsymbol{\mu}|^2 - |\mathbf{k} \cdot \boldsymbol{\mu}|^2$.  We will see, however, that the chain introduces an important modification to this pattern, which is due to the atom interacting with its nearest neighbours.  

Suppose that the system is initially prepared at $t=0$ in a single photon state ($M=1$).
We wish to calculate the photon emission probability per unit solid angle, $d \Gamma / d \Omega$, which gives the transition rate for a photon to be emitted in the direction $(\theta, \phi)$ per unit solid angle and for the atom chain to go into the ground state ($M=0$).  The matrix element of $\hat{V}$ for the system in the $M=1$ state to emit a photon of polarization $\boldsymbol{e}_{\mathbf{k}s}$ is given by
\begin{equation}
\bra{\psi^{1}_g} \hat{V} \ket{0} = - i \hbar
 \left( \frac{\omega_k}{2 \epsilon_0 \hbar V} \right)^{1/2}
\mathbf{e}_{\mathbf{k}s} \cdot \bra{\psi^{1}_g} \hat{\boldsymbol{\mu}} \ket{0},
\end{equation}
where
\begin{equation}
\hat{\boldsymbol{\mu}} = \boldsymbol{\mu}  \sum_{j=1}^{N} e^{-i \mathbf{k} \cdot \mathbf{r}_j} \hat{S}^+_j + \mathrm{H.c.}
\end{equation}
is the transition dipole moment operator and $\mathbf{e}_{\mathbf{k}s}$ is the polarisation vector of the emitted photon of wave vector $\mathbf{k}$ and polarisation $s$.  The vector $\mathbf{r}_i$ is the position of the $i$-th atom.   
We first calculate the transition dipole moment between the single-photon energy eigenstate and the ground state.  Using (\ref{eq:energyEigenstates}), we find
\begin{eqnarray}
\bra{\psi^{1}_{g}} \hat{\boldsymbol{\mu}} \ket{0} & = & \bra{\psi} \boldsymbol{\mu} \sum_{j=1}^{N} e^{-i \mathbf{k} \cdot \mathbf{r}_j} \ket{j} \nonumber \\
& = & \boldsymbol{\mu} \sum_{j=1}^{N} \sqrt{ \frac{2}{N+1} } e^{-i \mathbf{k}\cdot \mathbf{r}_j} \sin g j \xi .
\end{eqnarray}
By expressing the sine function in terms of complex exponentials, we find that there are two geometric summations generated by the sum over $j$.  Using the formula for the summation of a geometric series, this results in the equation
\begin{eqnarray}
\bra{\psi^{1}_g} \hat{\boldsymbol{\mu}} \ket{0} & = & \boldsymbol{\mu} \frac{e^{-i \mathbf{k} \cdot \mathbf{r}_1 + i \mathbf{k} \cdot \mathbf{a}}}{\sqrt{2N+2}} \cdot \left[ 1 - (-1)^g e^{-i(N+1) \mathbf{k} \cdot \mathbf{a}} \right] \cdot
\frac{\sin g\xi}{\cos \mathbf{k} \cdot \mathbf{a} - \cos g \xi},
\label{eq:geometric}
\end{eqnarray}
where $\mathbf{a} = \mathbf{r}_{i+1} - \mathbf{r}_i$ is the displacement between two adjacent atoms.
The transition probability per unit solid angle for a $M=1$ state to emit a photon of polarization $\boldsymbol{e}_{\mathbf{k}s}$ is then given by
\begin{eqnarray}
\frac{\rmd \Gamma_{g, \, s}}{\rmd \Omega} & = & \frac{2\pi}{\hbar} | \bra{\psi^1_g} \hat{V} \ket{0} |^2
\frac{V \omega^2}{(2 \pi c)^3 \hbar} \nonumber \\
& = &  \left( \frac{\omega_k^3 | \mathbf{e}_{\mathbf{k}s} \cdot \boldsymbol{\mu} |^2
}{8 \pi^2 c^3 \epsilon_0 \hbar} \right) 
\left( \frac{\sin^2 g \xi}{N+1} \right) 
\frac{1 - (-1)^g \cos [(N+1) \mathbf{k} \cdot \mathbf{a}]}{( \cos \mathbf{k} \cdot \mathbf{a} - \cos g \xi )^2}.
\end{eqnarray}

We now sum over the two polarizations $s$ of the radiation field.  By completeness, we have
\begin{equation}
|\mathbf{e}_{\mathbf{k}1} \cdot \boldsymbol{\mu} |^2 + |\mathbf{e}_{\mathbf{k}2} \cdot \boldsymbol{\mu} |^2 = |\boldsymbol{\mu}|^2 - |\boldsymbol{\mu} \cdot \mathbf{\hat{k}}|^2,
\end{equation}
where $\mathbf{\hat{k}}$ is the unit vector in the direction of the emitted radiation wave vector $\mathbf{k}$.  The total transition rate per unit solid angle in the far-field zone is then given by
\begin{eqnarray}
\frac{\rmd \Gamma_g}{\rmd \Omega} & = & \left( \frac{\omega_k^3 (| \boldsymbol{\mu} |^2 - | \mathbf{\hat{k}} \cdot \boldsymbol{\mu} |^2)
}{8 \pi^2 c^3 \epsilon_0 \hbar} \right) 
\left( \frac{\sin^2 g \xi}{N+1} \right)
\frac{1 - (-1)^g \cos [(N+1) \mathbf{k} \cdot \mathbf{a} ]}{( \cos \mathbf{k} \cdot \mathbf{a} - \cos g \xi )^2} \\
&=& \left( \frac{\omega_k^3 (| \boldsymbol{\mu} |^2 - | \mathbf{\hat{k}} \cdot \boldsymbol{\mu} |^2)
}{8 \pi^2 c^3 \epsilon_0 \hbar} \right) f(\mathbf{k}, g, N).
\label{eq:totalDifferentialDecay}
\end{eqnarray}
The factor in parentheses is the familiar spontaneous emission rate of one atom in isolation; the geometry of the chain and its multi-atom effects have been absorbed into the structure factor $f(\mathbf{k}, g, N)$:
\begin{equation}
f(\mathbf{k}, g, N) = \frac{1 - (-1)^g \cos [(N+1) \mathbf{k} \cdot \mathbf{a}]}{( \cos \mathbf{k} \cdot \mathbf{a} - \cos g \xi )^2}
\left( \frac{\sin^2 g \xi}{N+1} \right).
\label{eq:structureFactor}
\end{equation}
The structure factor $f(\mathbf{k}, g, N)$ provides a measure of the collective behaviour of the chain, due to the specific arrangement of the atoms.  We will focus on this object for the rest of the paper, to see what radiative effects are unique to a linear chain, and think of the structure factor as a ``radiation pattern.''  It is important to remember, however, that the actual radiation pattern will also be modulated by the sine of the angle between the emitted photon's wave vector, and the atomic dipole moment.

It can clearly be seen that the structure factor is symmetric about the centre of the chain, as evidenced by the fact that all the $\mathbf{k} \cdot \mathbf{a}$ terms are operated on by cosine functions.  There is also the interesting factor $1 - (-1)^g \cos [(N+1) \mathbf{k} \cdot \mathbf{a}]$ present in the equation.  In the small sample limit, this would become $1 - (-1)^g$, which would suggest that in such a limit all the $g$-even states would not decay, in agreement with our previous paper.  Thus $g$ being even is a \emph{sufficient} condition for an eigenstate to be subradiant in the small sample limit.  When $g$ is odd, the state may be subradiant or superradiant depending on the particular case.  We must keep in mind however that these classifications becomes more tenuous when phase differences between adjacent atoms become significant.

\begin{figure}
\begin{center}
\includegraphics{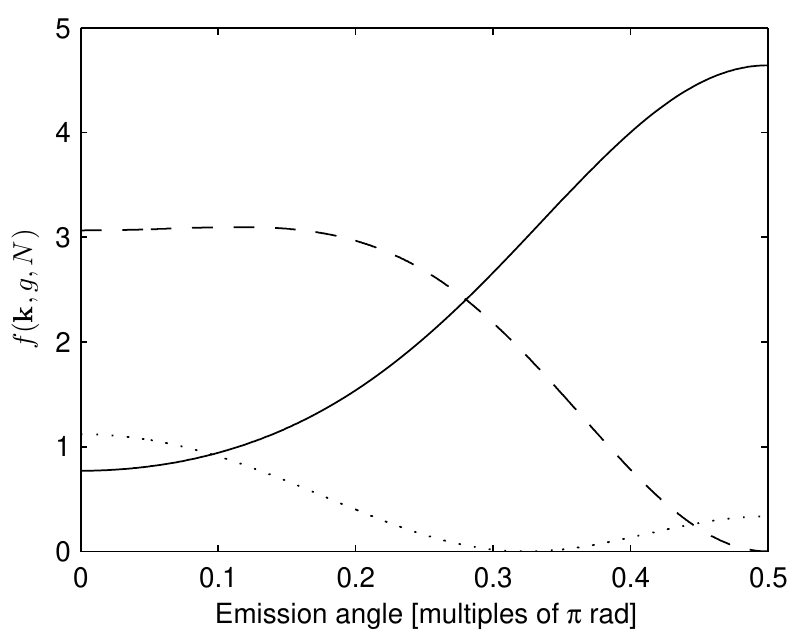}
\end{center}
\caption{Plot of the structure factor $f(\mathbf{k}, g, N)$ for a five-atom chain against observation direction $\mathbf{k}$ in radians for different values of $g$.  The cases plotted are for $g=1$ (solid line), $g=2$ (dashed line), and $g=3$ (dotted line).}
\label{fig:gvar}
\end{figure}

One may notice that the structure factor as given above becomes undefined when $\cos \mathbf{k} \cdot \mathbf{a} = \cos g \xi$.  However, this ambiguity is only an apparent one.  If we repeat the derivation of the detection rate formula, evaluating it before summing the geometric series, we obtain the following:
\begin{equation}
f(\mathbf{k}, g, N)|_{\cos \mathbf{k} \cdot \mathbf{a} = \cos g \xi} = \frac{N+1}{2},
\end{equation}
a result which is consistent with taking the appropriate limit of (\ref{eq:structureFactor}) at the undefined point.  A graph of the structure factor, (\ref{eq:structureFactor}), is plotted in figure \ref{fig:gvar} for a five-atom chain for a few $g$ values.  One can clearly see an overall decrease in the structure factor for higher values of $g$ up to $g=3$.  Also, we note that the radiation is not isotropic, being either focused along the length of the chain, or perpendicular to it according to the value of $g$.

The small sample behaviour of the structure factor is also interesting.  Taking the limit $\mathbf{k} \cdot \mathbf{a} \rightarrow 0$, we obtain the expression
\begin{eqnarray}
f(g,N) & = & \frac{1 - (-1)^g}{(1 - \cos g \xi)^2} \left( \frac{\sin^2 g \xi}{N+1} \right) \nonumber \\
& = & \frac{1 - (-1)^g}{N+1} \cot^2 \smallfrac{1}{2} g \xi.
\label{eq:smallSample}
\end{eqnarray}
The form of (\ref{eq:smallSample}) displays the radiation rate of closely spaced atoms.  It is zero for even values of $g$, independent of the number of atoms.  For odd values of $g$, the $\cot^2 \smallfrac{1}{2} g \xi$ term decreases as $g$ increases from $1$ to $\lceil N/2 \rceil$, and then increases from $\lceil N/2 \rceil$ to $N$.  For a large number of atoms, the maximum value the structure factor can attain is
\begin{equation}
f_{\mathrm{max}}(g,N) = \frac{8}{\pi^2} (N+1),
\end{equation}
while the lowest value for the structure factor is zero.

\subsection{Radiation from two and three atoms}

As an illustration of our general formula for the radiation pattern, consider a chain composed of $N=2$ atoms.  In this case, we can recover the known form of the structure factor for two atoms as given in the literature \cite{bib:Ficek2002, bib:Lehmberg1970}.  For $N=2$, the structure factor (\ref{eq:structureFactor}) simplifies to:

\begin{eqnarray}
f(\mathbf{k}, g, 2) = \frac{1 - (-1)^g \cos (3 \mathbf{k} \cdot \mathbf{a})}{( \cos \mathbf{k} \cdot \mathbf{a} +\smallfrac{1}{2}(-1)^g )^2}
\left( \frac{1}{4} \right).
\end{eqnarray}

Using the Chebyshev polynomial $T_3(x) = 4x^3 -3x$, we observe that $\cos 3 \alpha = 4 \cos^3 \alpha - 3 \cos \alpha$.  This allows us to simplify the above expression (by dividing the polynomial in $\cos \mathbf{k} \cdot \mathbf{a}$ in the denominator into the polynomial in the numerator) to
\begin{eqnarray}
f(\mathbf{k}, g, 2) = 1 - (-1)^g \cos \mathbf{k} \cdot \mathbf{a},
\end{eqnarray}
which reveals the radiation pattern of the symmetric ($g=1$) and antisymmetric ($g=2$) states of the two dipole-coupled atoms \cite{bib:Ficek2002, bib:Lehmberg1970}.  Taking the small sample limit ($\mathbf{k} \cdot \mathbf{a} \rightarrow 0$), we see that symmetric states decay twice as fast as an isolated atom, whilst antisymmetric states do not decay.

For comparison, the structure factor for $N=3$ is given by
\begin{eqnarray}
f(\mathbf{k}, g, 3) & = & 1 + 2c^1_g c^2_g [ 1 - (-1)^g ] \cos \mathbf{k} \cdot \mathbf{a} - 2 (c^1_g)^2 (-1)^g \cos 2 \mathbf{k} \cdot \mathbf{a},
\label{eq:threeatom}
\end{eqnarray}
where $c^k_g = \sqrt{2/(N+1)} \sin g k \xi$.  In the small sample limit of $ka \rightarrow 0$, the structure factor is independent of the wave vector $\mathbf{k}$ and thus becomes a direct measure of the total collective decay rate.  In particular, an eigenstate is superradiant or subradiant if $f>1$ or $f<1$, respectively.  We find that for the case of $N=3$, in the small sample limit of (\ref{eq:threeatom}), there are two subradiant states since  $f(g=2) = 0$ and $f(g=3) = 3/2 - \sqrt{2}$, and one superradiant state since $f(g=1) = 3/2+\sqrt{2}$.

\section{Radiation from one excited atom}
\label{sec:singleAtomDecay}
In the previous section we demonstrated the general radiative properties of a chain of collectively radiating qubits.  We examine here the radiation pattern that arises from the transition $\ket{j \,} \rightarrow \ket{0}$.  The state $\ket{j \,}$ corresponds to one atom in the chain being in the excited state whilst all the other atoms are in the ground state.  In considering such a transition, we can see how the dipole-dipole coupling, or more pictorially, the presence of the other atoms in the chain, affect the radiation pattern of a single atom through the structure factor.  The lowest order process that can occur is an initial state $\ket{j\,}$ at $t_1$ propagating to a time $t_3$, emitting a photon, and then continuing to propagate to a time $t_2$.  For clarity, we set $\hbar =1$ in this section.

\begin{figure}
\begin{center}
\includegraphics{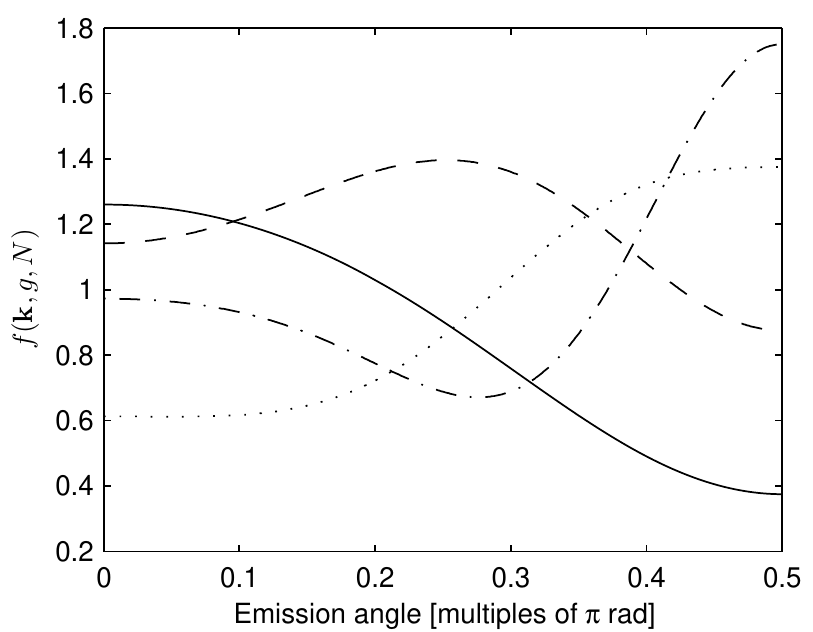}
\end{center}
\caption{Structure factor for a seven atom chain ($ka = 1$), for the initial states $\ket{1}$ (solid line), $\ket{2}$ (dashed line), $\ket{3}$ (dotted line), and $\ket{4}$ (dot-dash line).}
\label{fig:jvar}
\end{figure}

The propagator for the nonradiating atom chain is given by
\begin{equation}
K(2,1) = \sum_g e^{-iE_g(t_2 - t_1)} \ket{\psi^1_g} \bra{\psi^1_g}.
\end{equation}
The amplitude for the initial state to radiate a photon is given by
\begin{eqnarray}
\bra{0} K_V(2,1) \ket{j \,} & \equiv & i\int_{t_1}^{t_2} dt_3 \bra{0} K(2,3) V K(3,1) \ket{j \,} \nonumber\\
&=& i\sum_g c^j_g \bra{0} V \ket{\psi^1_g} 2\pi \delta(\omega - E_g),
\label{eq:eigenamp}
\end{eqnarray}
where we have set $t_2 = -t_1 = T/2$ and approximated the integral over $dt_3$ with a delta function.  We now switch to a discrete electromagnetic field spectrum to convert the Dirac delta function to a Kronecker delta.  This implies the replacement
\begin{equation}
\delta(\omega - E_g) \rightarrow \frac{T}{2 \pi} \delta_{\omega, E_g}.
\end{equation}
To remain consistent, any integral over $\omega$ must be replaced by a discrete sum of the form
\begin{equation}
\int d \omega \rightarrow \frac{2 \pi}{T} \sum_{\omega}.
\end{equation}
The expression on the right-hand side of (\ref{eq:eigenamp}) becomes
\begin{equation}
i\sum_g c^j_g \bra{0} V \ket{\psi^1_g} T \delta_{\omega, E_g}.
\end{equation}
We now square this expression to obtain a transition probability, and sum over all possible emitted photon frequencies $\omega$, weighted by the density of final states, to obtain the total transition probability per unit solid angle:
\begin{eqnarray}
\frac{\rmd \Gamma^j}{\rmd \Omega} T & =& \frac{2 \pi}{T} \sum_{\omega} \sum_{g,g'} c^j_g c^j_{g'} \bra{\psi^1_g} V \ket{0} \bra{0} V \ket{\psi^1_g} T^2 \delta_{\omega, E_g} \delta_{\omega, E_{g'}} \rho(\omega).
\end{eqnarray}
Clearly, the sum over $\omega$ forces $E_g = E_{g'}$ through the product of the two Kronecker deltas.  Summing over $g'$, we finally obtain
\begin{eqnarray}
\frac{\rmd \Gamma^j}{\rmd \Omega} & = & 2 \pi \sum_g (c^j_g)^2 |\bra{\psi^1_g} V \ket{0}|^2 \rho(E_g) \nonumber \\
& = & \sum_g (c^j_g)^2 \frac{\rmd \Gamma_g}{\rmd \Omega}.
\end{eqnarray}
The radiation pattern from several different initial states for a seven-atom chain is shown in figure \ref{fig:jvar}.  Naively, we may expect two different radiation patterns in general: (a) the situation where the excited atom is at either end of the chain, and hence has only one interacting neighbor, and (b) the excited atom at any other position in the chain, surrounded by two interacting neighbors.  However, as can be seen from the graph, this is not the case.  This is probably due to the fact that the eigenstate excitation amplitudes (i.e. the $c^j_g$) vary along the chain.  Given the reflection symmetry of the chain, one instead classifies the patterns into $\lceil N/2 \rceil$ groups, since the $j$-th and $(N+1-j)$-th atoms should radiate the same way by symmetry.  Thus, the radiation pattern from the system depends in an essential manner on the initial excitation of the system.

\begin{figure}
\begin{center}
\includegraphics{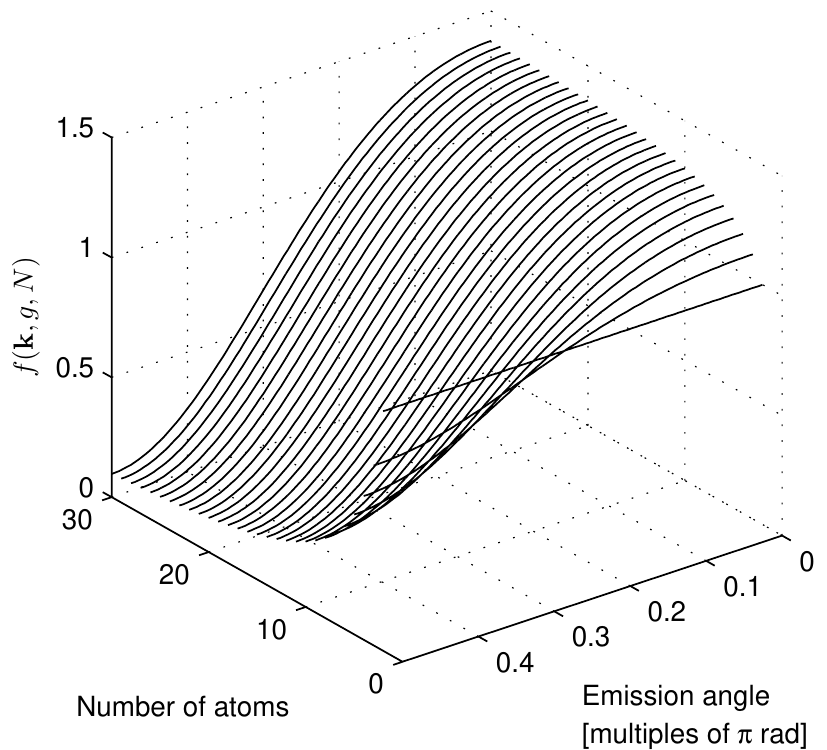}
\end{center}
\caption{Structure factor profile of the state $\ket{1}$ as a function of the number of atoms and angle of emission relative to the interatomic axis with $ka = 1$.}
\label{fig:nvar}
\end{figure}

It is also interesting to see how the radiation pattern varies as a function of the number of atoms.  This is shown in figure \ref{fig:nvar}, where the initial state is $\ket{1}$ (one atom at the end of the chain excited, all others in the ground state), and the distance between adjacent atoms is held constant.  We can see an interesting modification in the directionality of the radiation pattern with the number of atoms.  Whilst the radiation pattern is spherically symmetric for two atoms, it becomes more concentrated along the axis of the chain.

\section{Diffraction properties of the structure factor}
In the preceding sections, we looked at how the structure factor shapes the radiation field for two different states of the chain.  We now focus on general properties of the structure factor itself and how, by altering the number of atoms in the chain and the distances between them, one may alter the directional properties of the radiation field emitted by spontaneous emission.

From (\ref{eq:structureFactor}), we may expect to see effects in the radiation distribution similar to $N$-slit diffraction.  Using trigonometric identities, the structure factor can be expressed in the following form:

\begin{eqnarray}
f(\mathbf{k}, g, N) &=& \frac{\sin^2 \smallfrac{1}{2}(N+1)(\mathbf{k} \cdot \mathbf{a} - g \xi)}{2 \sin^2 \smallfrac{1}{2}(\mathbf{k} \cdot \mathbf{a} + g \xi) \sin^2 \smallfrac{1}{2}(\mathbf{k} \cdot \mathbf{a} - g \xi)}
\left( \frac{\sin^2 g \xi}{N+1} \right)
\end{eqnarray}

We see the the structure factor contains the same modulating factor type as exists in the diffraction of $N+1$ slits \cite{bib:Klein1986}, namely
\begin{equation}
\sin^2 [ \smallfrac{1}{2}(N+1) (\mathbf{k} \cdot \mathbf{a} - g \xi) ] / \sin^2 [ \smallfrac{1}{2} (\mathbf{k} \cdot \mathbf{a} - g \xi)],
\end{equation}
where the factor $(\mathbf{k} \cdot \mathbf{a} - g \xi)$ is analogous to the phase difference between two adjacent slits.

\begin{figure}
\begin{center}
\includegraphics{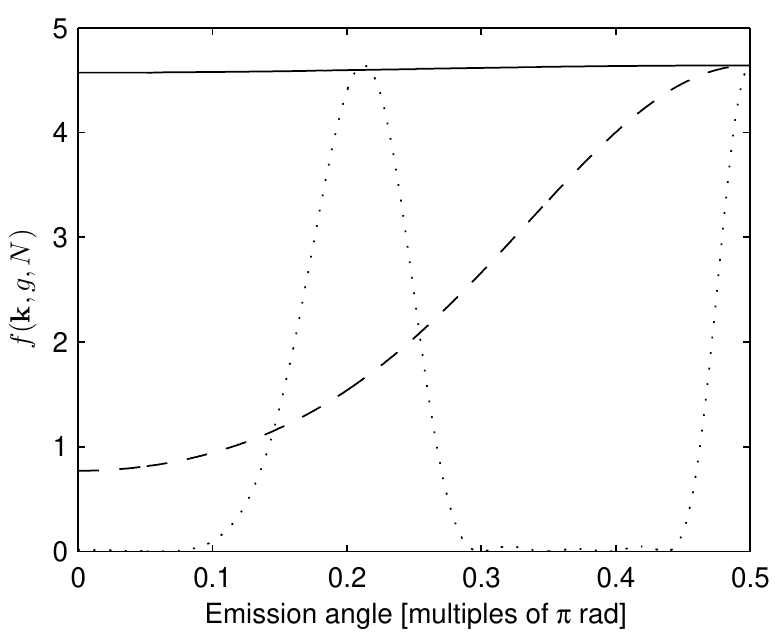}
\end{center}
\caption{\label{fig:krvar} Plot of the structure factor $f(\mathbf{k}, g=1, N)$ for a five-atom chain against observation direction $\mathbf{k}$ in radians for different values of $ka$.  The cases plotted are $ka=0.1$ (solid line), $ka=1$ (dashed line), and $ka=8$ (dotted line).}
\end{figure}

We therefore expect that any diffraction maxima get narrower and more higher-order peaks are introduced as more atoms are placed in the chain.  This can be seen if we plot the structure factor for a few different values of $ka$, as shown in figure \ref{fig:krvar}.

\begin{figure}
\begin{center}
\includegraphics{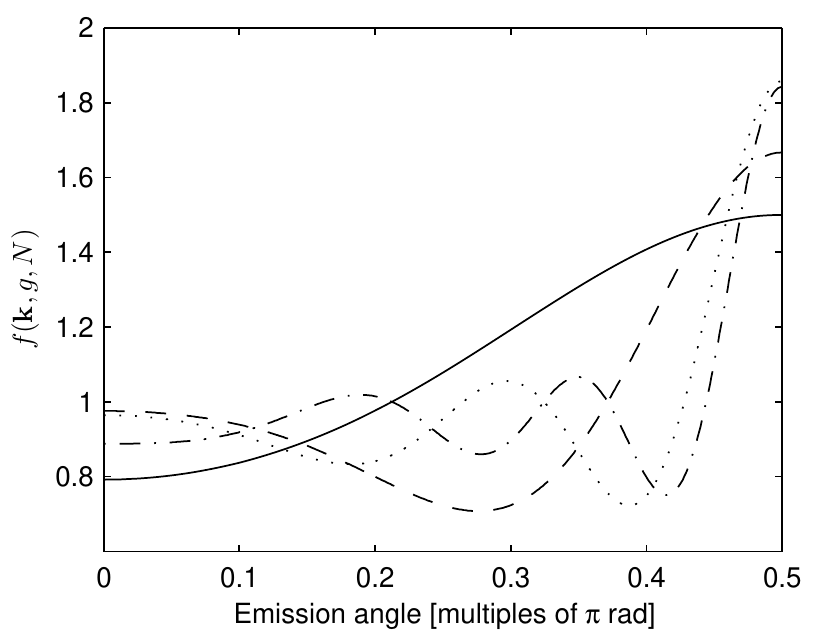}
\end{center}
\caption{\label{fig:nvar_jcentral_2d} Structure factor of the state $\ket{\lceil N/2 \rceil}$ for different values of $N$, $ka = 0.1$.  The cases plotted are for $N=3$ (solid line), $N=8$ (dashed line), $N=13$ (dotted line), $N=18$ (dot-dashed line).}
\end{figure}

\begin{figure}
\begin{center}
\includegraphics{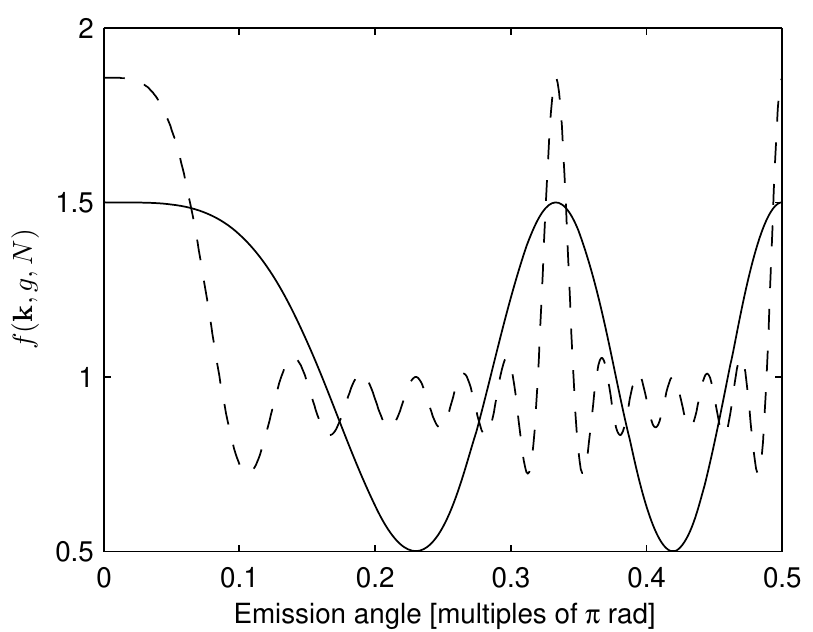}
\end{center}
\caption{\label{fig:nvar_jcentral_2d_kr2pi} Structure factor of the state $\ket{\lceil N/2 \rceil}$ for different values of $N$, $ka = 2\pi$.  The cases plotted are for $N=3$ (solid line) and $N=13$ (dashed line).}
\end{figure}

\begin{figure}
\begin{center}
\includegraphics{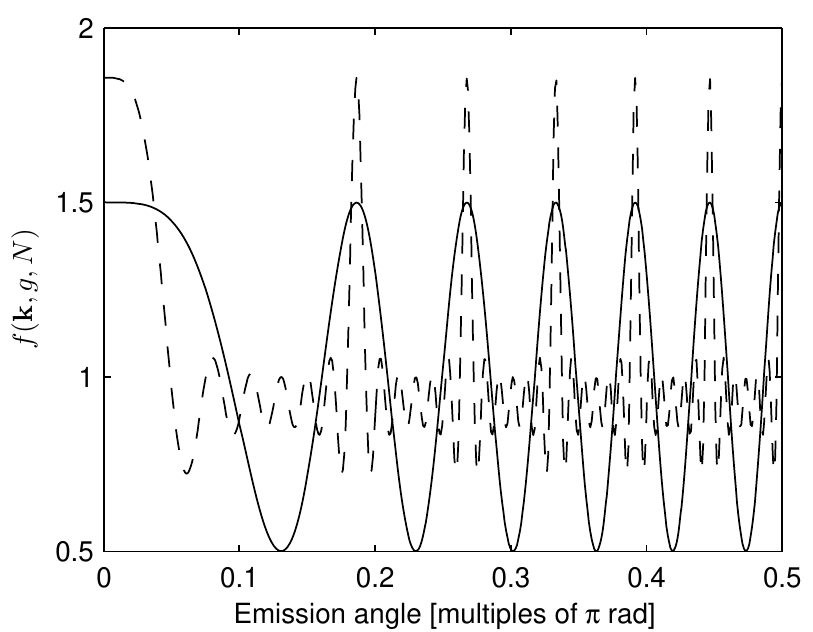}
\end{center}
\caption{\label{fig:nvar_jcentral_2d_kr6pi} Structure factor of the state $\ket{\lceil N/2 \rceil}$ for different values of $N$, $ka = 6\pi$.  The cases plotted are for $N=3$ (solid line) and $N=13$ (dashed line).}
\end{figure}

We can also analyse diffraction-like effects from single-atom radiation, shown in Figures \ref{fig:nvar_jcentral_2d}, \ref{fig:nvar_jcentral_2d_kr2pi}, and \ref{fig:nvar_jcentral_2d_kr6pi}.  In these graphs, there are several peaks which do not change as the number of atoms in the chain is increased.  These are analogous to the the principle maxima for the diffraction of $(N+1)$ slits.  We see that as more atoms are added to the chain, the positions of these maxima do not change, and that an increasing number of smaller maxima are added in between, consistent with our analogy of multiple-slit diffraction.  All peaks do get narrower as the number of atoms is increased, which is consistent with our diffraction interpretation.

Clearly, the structure factor profile is punctuated by peaks which become increasingly sharp as more atoms are added to the chain.  In addition, reducing $ka$ decreases the number of peaks.  Therefore, to make the emitted radiation more directional along the interatomic axis, one simply reduces $ka$ (or more physically, the distance between the atoms), and increases the number of atoms.   When $ka > 2\pi$, one would expect repetitions in the  angular profile of the structure factor.  Thus, $ka$ should be roughly $2\pi$.  If it is significantly less than this, one starts to lose most of the interesting features of the structure factor, such as the sharp peaks shown in several previous figures.


\section{Total decay rate}
\label{sec:totalDecayRate}
We can also find the total decay rate of a particular $M=1$ eigenstate by integrating over all the possible directions of the wave vector $\mathbf{k}$.  This integration can be simplified by re-deriving the total differential decay rate (\ref{eq:totalDifferentialDecay}) without step (\ref{eq:geometric}), the original purpose of which removes the summations in the expression, but makes integration harder.  Following this modified procedure, we arrive at an alternate expression for the differential decay rate:
\begin{equation}
\frac{\rmd \Gamma_g}{\rmd \Omega} = \left( \frac{\omega_k^3 (| \boldsymbol{\mu} |^2 - | \mathbf{\hat{k}} \cdot \boldsymbol{\mu} |^2)
}{8 \pi^2 c^3 \epsilon_0 \hbar} \right)
\sum_{p, \, q=1}^N e^{-i \mathbf{k} \cdot \mathbf{a} (p-q)} c^p_g c^q_g.
\end{equation}
Integrating over all directions gives the result
\begin{equation}
\Gamma_g = \gamma + \gamma \sum_{ \substack{i,j=1\cr i \neq j} }^N F(k r_{ij}) c^i_g c^j_g,
\end{equation}
where $\gamma$ is the decay rate for a single atom in isolation and
\begin{eqnarray}
&& \fl F(kr_{ij}) = \frac{3}{2} \left\{ [ 1-(\boldsymbol{\hat{\mu}} \cdot \mathbf{\hat{r}}_{ij})^2 ] \frac{\sin kr_{ij}}{kr_{ij}}
+ [1-3(\boldsymbol{\hat{\mu}} \cdot \mathbf{\hat{r}}_{ij})^2] \left[\frac{\cos kr_{ij}}{(kr_{ij})^2} - \frac{\sin kr_{ij}}{(kr_{ij})^3} \right] \right\}.
\end{eqnarray}
We have defined $\mathbf{\hat{r}}_{ij} \equiv \mathbf{r}_{ij}/r_{ij}$, $\mathbf{r}_{ij} \equiv \mathbf{r}_j - \mathbf{r}_i$, and $r_{ij} \equiv |\mathbf{r}_{ij}|$.

\begin{figure}
\begin{center}
\includegraphics{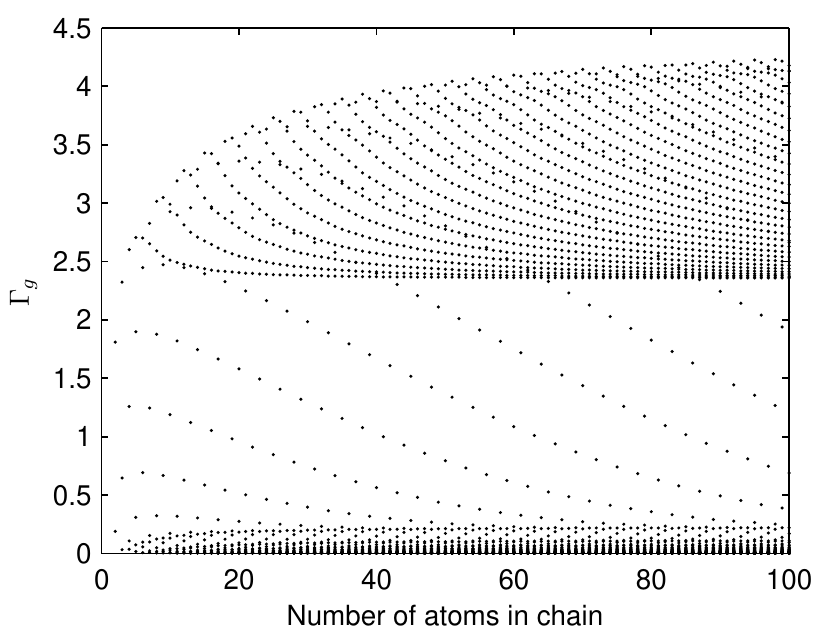}
\end{center}
\caption{\label{fig:allrates} The decay rates of all eigenstates plotted against the number of atoms in the chain for the case $\boldsymbol{\mu} \cdot \mathbf{a} = 0$ and $\mathbf{k} \cdot \mathbf{a} = 1$.  Note that for this graph $\gamma=1$.}
\end{figure}

We would like to get a sense of consequences of this equation.  For example, some might expect that the collective interaction of the atoms through the vacuum field hastens decoherence of the system.  Let us explore such a suggestion here.

Figure \ref{fig:allrates} shows a plot of the decay rates for all $M=1$ eigenstates as a function of the number of atoms.  This plot is to be understood as follows: there are $N$ eigenstates in the $M=1$ subspace, where $N$ is the number of atoms in the chain.  These states are represented by $\ket{\psi_{g}^1}$, where $1 \le g \le N$; the decay rate of a particular eigenstate $\ket{\psi_{g}^1}$ is given by $\Gamma_{g}$.  Thus, for a chain with $N$ atoms, we will have $N$ decay rates $\Gamma_{g}$ for the $M=1$ subspace.  Thus in the plot shown, there are $N$ points associated with each value of $N$.

  We see that there is an interesting ``banding'' feature in the decay rates; the main features are an upper cluster of states or ``band,'' a lower cluster of states, and a small number of states interspersed between the two clusters.  The thickness of the ``bands'' varies with the orientation of the atom dipoles with respect to the chain axis.  We can distinguish three groups according to the rate at which they decay: superradiant ($\Gamma_g > \gamma$), natural ($\Gamma_g = \gamma$) and subradiant ($\Gamma_g < \gamma$).  From these definitions, we may classify the upper and lower bands as superradiant and subradiant bands.  The subradiant band is the most interesting feature in Figure \ref{fig:allrates}.  Since the eigenstates in this region decay slowly, such states have potential applications in regards to the storage of quantum information and photon trapping.  

\begin{figure}
\begin{center}
\includegraphics{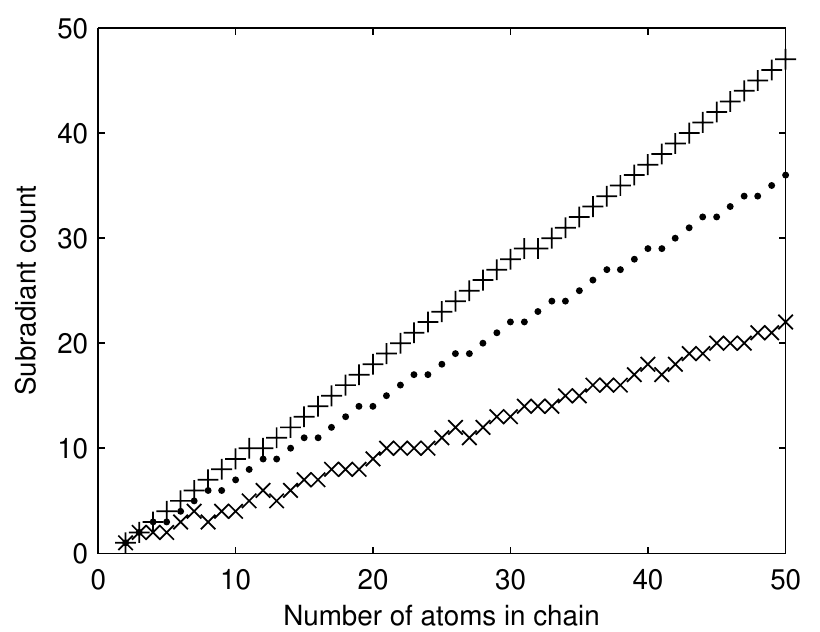} 
\end{center}
\caption{\label{fig:subradiant_count_ma1_3plots} Subradiant state count as a function of the number of atoms for the following cases: $ka=4.4$ (marked by $\times$'s), $ka=1$ (marked by $\cdot$'s) and the small sample limit $ka=0$ (marked by $+$'s).  We have set $\boldsymbol{\mu} \cdot \mathbf{a} = 1$.}
\end{figure}

To explore the size of the subradiant subspace, we consider the number of subradiant states against $N$, using the previously mentioned definition $\Gamma_g < \gamma$ for such states.  Figure \ref{fig:subradiant_count_ma1_3plots} reveals that the number of subradiant states varies linearly with the number of atoms in an approximate fashion.  For the few values of $ka$ considered in the graph, the linearity of the relationship is maintained.  Indeed, $ka$ appears to have the general feature of only altering the gradient of the line of best fit, as can be seen in the graph.  It is important to remember that we are dealing with a \emph{subspace} of eigenstates, so Dicke's result\cite{bib:Dicke1954} that half of the eigenstates are subradiant does not apply since that result is meant for the complete space of eigenstates.

Now imagine lines of best fit which run through the roughly linear plots in Figure \ref{fig:subradiant_count_ma1_3plots}.  The gradient of such a line can be interpreted as a rough estimate of the fraction of the number of eigenstates which are subradiant; let's say that the gradient of the best fit line was 0.42 -- then roughly 42\% of the eigenstates are subradiant.  This simple relationship is due to the fact that the line passes close to the origin, so we don't have to worry about a ``non-zero $y$-intercept.''  Such a quantity is useful since it allows us to make general statements about eigenstate lifetime \emph{independent} of the number of atoms in the chain, a consequence of the linear relationship.  Obviously, such an approximation only works only when there are more than a few atoms in the chain.

Let's now take a closer look to see how one can define a line of best fit.  Plots like the ones shown in Figure \ref{fig:subradiant_count_ma1_3plots} have the general form of straight lines with kinks in them to give a stepped appearance.  We can compute the gradient by taking the average of the slope of the lines connecting adjacent points.  This averaging procedure gives more weight to the straight lines that appear in the plots rather than the disjointed ``steps'' that connect them.  Such an averaging procedure gives a good line of best fit for the ranges of $ka$ and $N$ considered here.  The uncertainty in predicting the number of subradiant eigenstates from the gradient is of the order of a few states, but adequately serves the purpose of being a convenient semi-quantitative estimator for the size of the subradiant subspace for a linear chain.

We now have a quantity which is independent of the number of atoms in the chain: the gradient of the subradiant count, or ``subradiant fraction'' (i.e. ``the fraction of the total number of eigenstates which are subradiant'') to be more concise.  Let us have a look at how altering the only other major property of the system, $ka$, affects this quantity.

\begin{figure}
\begin{center}
\includegraphics{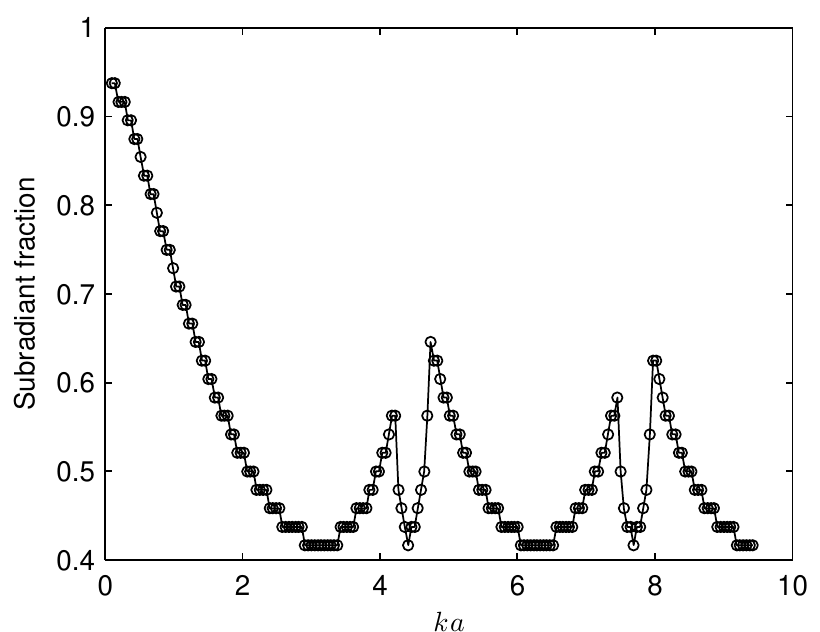} 
\caption{\label{fig:subradiant_fraction_ma_1_long_ka} Plot of averaged gradient or ``subradiant fraction'' against $ka$ for $\boldsymbol{\mu} \cdot \mathbf{a} = 1$.}
\end{center}
\end{figure}

A plot of the subradiant fraction (i.e. the ``gradient of the best-fit line to the subradiant count,'' the ``fraction of the total number of states which are subradiant'') against $ka$ is shown in figure \ref{fig:subradiant_fraction_ma_1_long_ka}.  One can see that for small interatomic separations such that $ka < 2$, more than half of the eigenstates are subradiant, approaching 94\% in the small sample limit.  As $ka$ increases beyond 2, we see a periodic pattern with interesting valley features where the number of subradiant states drops below 50 percent.

The fact that most of the eigenstates are subradiant for $ka < 2$ is a beneficial property of this system.  For example, to store quantum information, we need states that do not change in time for as long as possible.  The ideal state would be one that does not decay.  A collection of such states in which one can process quantum information is known as a \emph{decoherence-free subspace}.  Although this system does not have such a decoherence-free subspace, its \emph{subradiant subspace} is very large, and such a collection of states is still useful for quantum information, given that the other approaches to quantum computing \cite{bib:SchmidtKaler2003, bib:Cirac1995, bib:Steane1997} do not rely on collective behaviour and hence are limited by the natural decay rate $\gamma$.

Finally, we comment about the problem of individual addressing of the atoms in a possible realisation of single-qubit operations.  In the case of direct individual excitation and addressing of the atoms, this may be realised in a microcavity situation, where atoms can be kept at large distances that are still comparable to the resonant wavelength \cite{bib:osnaghi2001}.  Qubits in our system can be manipulated collectively, which clearly follows from the collective nature of the energy eigenstates.  Regarding the problem of the encoding of quantum information given that targeting a specific qubit requires resolution of the order of a resonant wavelength, a potential fix for this problem is to introduce a Stark shift to increase the spacing between the two energy levels in each qubit such that diffraction no longer presents itself as a problem.  This would also eliminate the dipole-dipole interaction, which is useful when direct single-qubit measurements are required.

\section{Conclusions}
We have studied the radiative properties of a chain of dipole-dipole interacting atoms.  Assuming that only one photon is present in the system, we have calculated the component of the radiation emission pattern corresponding to collective interactions, and also the total decay rates to study the decoherence property of the chain.

We were particularly interested in the directional distribution of the radiated field and the presence of the subradiant states that decay slower than a single atom in isolation.  We have shown that the radiation tends to be focused either along or perpendicular to the chain, and that in some ways the emission pattern is analogous to multiple-slit diffraction.  The total decay rates were also explored to see how many states are subradiant.  We found that the fraction of eigenstates that are subradiant depends on the distance between the atoms.  For small interatomic distances, we found that the majority of eigenstates are subradiant.

\ack
This research was supported by the Australian Research Council.

\end{document}